\documentclass[%
 reprint, 
nofootinbib,
pra,
]{revtex4-2}

\usepackage{graphicx}
\usepackage{dcolumn}
\usepackage{bm}

\usepackage{color}
\usepackage{xcolor}

\usepackage[english
]{babel}

\begin{document}

\title{\textbf{Role of plasma waves in rescattering processes in intense laser fields}}%
\author{
V.~V.~Strelkov$^{1\ast}$, S.~A.~Bondarenko$^{1,2}$, I. V. Smetanin$^{1}$
}
\affiliation{
\mbox{$^{1}$P.N. Lebedev Physical Institute of the Russian Academy of Sciences, 53 Leninskiy Prospekt,  Moscow 119991, Russia} \\
\mbox{$^{2}$National Research Nuclear University MEPhI, 31 Kashirskoe Highway, Moscow 115409, Russia} \\
$^{\ast}$strelkov.v@gmail.com
}

\date{\today}

\begin{abstract}
 Rescattering of the photoelectron at its parent ion underlie a number of phenomena in intense laser field interaction with matter, such as high harmonic generation, attosecond pulse production, non-sequential double ionization, and others. These processes are unavoidably accompanied by the medium photoionization. The interaction of the laser pulse with the photoionization-induced plasma  excites wakefield waves, which are self-consistently coupled plasma density and Langmuir waves. We study theoretically the effect of the electric field of the plasma wave on the rescattering processes. We show that this field can compensate for the magnetic drift of the rescattering electron, which otherwise dramatically suppresses the rescattering efficiency in intense low-frequency laser fields. Moreover, the presence of the plasma wave field leads to new lines in the spectrum emitted due to the XUV free-induction decay (XFID). Observation of these lines can allow, in particular, the detection of
forbidden transition frequencies, thus providing new perspectives for XFID spectroscopy. 
\end{abstract}

\maketitle


\section*{Introduction}

Rescattering processes taking place when intense femtosecond laser pulses interact with atoms, ions, and molecules, are currently among the key issues in laser physics and nonlinear optics. These processes are a cornerstone of attosecond pulse production~\cite{Marangos_2025, Ryabikin2023}, high-order harmonic generation (HHG)~\cite{Agostini_Nobel, Huillier_Nobel, Chang_2020}, non-sequential double ionization (NSDI)~\cite{Huillier_1982, Walker_1994, Becker_2012},  recollisional excitation~\cite{Hart_2000, Yudin_2001, Haan_2008}, high-energy above-threshold ionization (ATI) plateau~\cite{Paulus_1994} and other phenomena. The mechanism of the rescattering can be understood within a three-step semiclassical model~\cite{3-step_C, 3-step_S, 3-step_K} in which the elementary acts are: (i) ejection of an electron from an atom via tunneling ionization; (ii) its acceleration by the optical field which pushes the electron first away from the parent ion and then back toward it as the field changes sign, and (iii) the recollision of the electron with the parent ion.

Moreover, excitation by the laser pulse (in particular, the recollisional excitation) leads to an appearance of the atom or the ion in a coherent superposition of states. This superposition relaxes by coherently emitting XUV radiation [called XUV free induction decay (XFID)]. The excitation can occur via single photon~\cite{Cao2016, Bengtsson_2017} or multiphoton~\cite{Beaulieu_2017, Bloch_2019} absorption.

However, the propagation of the intense laser pulse in the medium also causes collective motion of the particles (first of all, plasma dynamics), which provides additional fields in the medium.  Their effect on the recollision processes is the scope of this paper.

The interaction of the laser pulse with a gas typically leads to a partial ionization of the medium. Plasma is produced in front of the main pulse due to the intense pulse pedestal, and/or within the pulse itself as a result of various multiphoton or above-threshold ionization processes~\cite{Sm1}. The interaction of the pulse with the underdense plasma produces wakefield waves, which are self-consistently coupled plasma density and electrostatic Langmuir waves. The high-field physics community has focused intensive experimental and theoretical studies on wakefields since Tajima and Dawson proposed using laser radiation to excite plasma waves for electron acceleration~\cite{Sm2}. Laser wakefield acceleration (LWFA) is of great interest due to its ability to sustain extremely large acceleration gradients, which are several orders of magnitude larger than those of conventional radio-frequency linacs~\cite{Sm3, Sm4, Sm5}. However, such high electrostatic wakefield strengths are only attainable at relativistic intensities of the driving laser pulse. At moderate intensities, the excited wakefield amplitude is not as high as in LWFA, but it can significantly affect the nonlinear dynamics of the excited electrons in combined laser and electrostatic fields, thereby modifying the rescattering processes. 

The physical mechanism by which the laser pulse propagating in an underdense plasma self-consistently excites the Langmuir electrostatic field and space charge density oscillations is straightforward and based on the ponderomotive interaction, as has been well established since the 1950s~\cite{Sm6}. The ponderomotive force associated with the laser pulse envelope expels the electrons from the laser pulse region. The resulting charge separation creates a self-consistent electrostatic field that forces the electrons to move back, providing plasma density oscillations. When the characteristic length scale of the axial gradient of the laser pulse intensity profile is close to the plasma wavelength, the ponderomotive force can excite a large-amplitude~\cite{Uryupin_2016} plasma wave (wakefield) whose phase velocity is approximately equal to the group velocity of the laser pulse~\cite{Sm7}. The amplitude of the longitudinal electrostatic field is limited by wave breaking, which occurs when the maximum electron oscillation velocity becomes as large as the phase velocity of the wake wave~\cite{Sm6, Sm8, Sm9}. Note that in this highly non-linear regime of the interaction of the laser field with plasma high-harmonic generation via non-rescattering mechanisms takes place~\cite{Uryupin_2002}. At moderate laser pulse intensities, a regular nonlinear wake wave is formed behind the laser pulse, with an amplitude proportional to the peak intensity of the driving pulse~\cite{Sm10, Sm11}.

The paper is organized as follows: in the first section we study analytically the properties of the plasma wave depending on the laser pulse and the plasma parameters;  in the second section we find conditions when the field of the plasma wave can compensate for the magnetic drift of the recolliding electron; in the third section we simulate the XFID process in the presence of the plasma wave via solving numerically the time-dependent Scr\"odinger equation (TDSE) for an atom emitting XUV.

\section{Excitation of a plasma wave by a laser pulse}

\subsection{\label{sec_plasma_analyt} Plasma hydrodynamics}

We study the excitation of a longitudinal plasma wave propagating in the same direction that the laser field does. The 1D geometry is assumed. This approximation is reasonable near the beam axis and for a short femtosecond laser pulse, in which the longitudinal gradient of the laser intensity (and thus the gradient of the ponderomotive energy) is much higher than the transverse one. \textcolor{black}{We assume the conditions typical for the experiments studying the rescattering processes, namely the medium densities $10^{17}-10^{19}$~cm$^{-3}$ and the laser intensities $10^{14}-10^{15}$~W/cm$^2$ (thus well below the relativistic one).} We start from the 1D plasma hydrodynamics equations~\cite{Bulanov2001} (in atomic units):
\begin{equation}
\partial_t n_e + \partial_x \left( n_e v_{||}\right) = w (I),
\label{n_e}    
\end{equation}
\begin{equation}
\partial_t v_{||} + v_{||}\partial_x v_{||} = - E_s-\frac{\nabla E^2}{4 \omega_0^2}-\nu_c v_{||},
\label{v_||}    
\end{equation}
\begin{equation}
\partial_x E_{s} = 4 \pi (n_i-n_e),
\label{E_s}    
\end{equation}
\begin{equation}
n_i(x,t) = \int^t w(I(x,t'))dt',
\label{n_i}    
\end{equation}
where $n_i(x,t)$ and $n_e(x,t)$ are the ion and the electron densities, respectively(writing Eq.~(\ref{n_i}) we assume that ions do not move),  $w(I)$ is the ionization rate, $I$ is the laser intensity, $v_{||}$ is the electrons' \textcolor{black}{longitudinal} velocity, $E$ is the laser field amplitude, $\nu_c$ is the collision frequency, $E_s$ is the self-consistent longitudinal field of the plasma wave.

\textcolor{black}{For the considered laser intensities and plasma densities one can} assume that the charge density $\rho=n_e-n_i$, as well as $v_{||}$ and $E_s$ are small perturbations. \textcolor{black}{(In particular, this means that in Eq.~(\ref{n_e}) one can replace $n_e v_\parallel$ with $n_i v_\parallel$  omitting terms of higher order of  smallness.)} From Eq.~(\ref{n_e})-(\ref{n_i}) we have:
\begin{equation}
    \partial_t \rho + \partial_x (n_i v_{||})=0
    \label{d_rho}
\end{equation}
\begin{equation}
    \partial_t v_{||} = - E_s -\frac{\nabla E^2}{4 \omega_0^2} - \nu_c v_{||}
    \label{d_v}
\end{equation}
\begin{equation}
    \partial_x E_s = - 4 \pi \rho
    \label{d_E_s}
\end{equation}

Differentiating Eq.~(\ref{d_v}) with respect to $t$ we have: 
\begin{equation}
    \partial_{tt}^2 v_{||} = - \partial_t E_s -\partial_t \Pi  - \nu_c \partial_t v_{||}
    \label{d2_v}
\end{equation}
where $\Pi=\frac{\nabla E^2}{4 \omega_0^2}$ is the ponderomotive \textcolor{black}{force}.

Differentiating Eq.~(\ref{d_E_s}) with respect to $t$ we have:
\begin{equation}
    \partial_{tx}^2 E_s = - 4 \pi \partial_t \rho,
    \label{d2_E_s}
\end{equation}
and taking into account Eq.~(\ref{d_rho}) we have from Eq.~(\ref{d2_E_s})
\begin{equation}
    \partial_{tx}^2 E_s = 4 \pi \partial_x (n_i v_{||}).
    \label{dtx_E_s}
\end{equation}
Integrating the latter Eq. over $x$ we have 
\begin{equation}
    \partial_{t} E_s = 4 \pi n_i v_{||}+const.
    \label{dt_E_s}
\end{equation}
Taking into account that $E_s \rightarrow 0$ for $x \rightarrow \pm \infty $ we find that $const=0$.

From Eqs.~(\ref{d2_v}) and~(\ref{dt_E_s}) we have
\begin{equation}
    \partial_{tt}^2 v_{||} +4 \pi n_i  v_{||} + \nu_c \partial_t v_{||} = -\partial_t \Pi 
    \label{dtt_v}
\end{equation}
Finding $v_{||}$ from this Eq. one can find $E_s$ from Eq.~(\ref{dt_E_s}) and $\rho$ from Eq.~(\ref{d_E_s}).

Below we consider a Gaussian laser pulse:
\begin{equation}
  E(t,x)=E_0 \exp{\left(-\frac{(t-x/c)^2}{\tau^2}\right)}. 
  \label{pulse}
\end{equation}

We assume that its peak amplitude $E_0$ does not depend on $x$; such approximation is reasonable near the focus of the laser beam.

\subsection{Plasma exists before the laser pulse ($n_i = const)$}

Here we consider the case when the plasma with density $n_i$ exists before the laser pulse and the pulse further ionizes the medium just slightly, so we assume $n_i=const$. In particular, this is the case of HHG in a plasma plume (see~\cite{Ganeev2013, Singh2021} and references therein). In this case we have from Eq.~(\ref{d_v}) and~(\ref{dt_E_s}) for $x=0$:
\begin{equation}
\partial^2_{tt}E_s+\omega_p^2 E_s+\nu_c \partial_t E_s=-\frac{E_0^2 t \omega_p^2} {\omega_0^2 \tau^2 c} \exp{\left(-\frac{2 t^2}{\tau^2}\right)}.
    \label{dtt_E_s}
\end{equation}
where $\omega_p=\sqrt{4 \pi n_i}$ is the plasma frequency. Introducing $$\theta=t \omega_p,$$ 
\begin{equation}
D=\tau \omega_p,
\label{D}
\end{equation}
we rewrite the latter Eq. as:

\begin{equation}
\partial^2_{\theta \theta}E_s+ E_s+\frac{\nu_c}{\omega_p} \partial_\theta E_s=- A \frac{\theta } {D^2} \exp{\left(-\frac{2 \theta^2}{D^2}\right)},
    \label{dthth_E_s}
\end{equation}
where 
\begin{equation}
A=\frac{E_0^2 \omega_p}{c \omega_0^2 }.
    \label{A}
\end{equation}

\subsection{Plasma appears during the laser pulse ($n_i \ne const$)} 
Here we consider the case where the laser pulse propagates in a gas, so initially $n_i=0$ and $n_i$ increases because of the gas photoionization. In this case we find from Eq.~(\ref{dt_E_s})

\begin{equation}
    \partial^2_{tt} E_s = 4 \pi n_i \partial_{t}v_{||}+4\pi v_{||}\partial_{t} n_i, 
 \end{equation}    
substituting $\partial_{t}v_{||}$ from Eq.~(\ref{d_v}) we have:    
\begin{equation}
     \partial^2_{tt} E_s = 4 \pi n_i (- E_s -\frac{\nabla E^2}{4 \omega_0^2} - \nu_c v_{||}) +4 \pi v_{||} \partial_{t} n_i. 
  \end{equation} 
One can rewrite it in the following form
\begin{equation}
    \partial^2_{tt} E_s +4 \pi n_i E_s +4 \pi n_i v_{||} (\nu_c- \partial_{t} n_i/n_i ) = -4 \pi n_i \frac{\nabla E^2}{4 \omega_0^2} .
\end{equation}
Using Eq.~(\ref{dt_E_s}) again, we find
\begin{equation}
    \partial^2_{tt} E_s +4 \pi n_i E_s + (\nu_c- \partial_{t} n_i/n_i ) \partial_t E_s = -4 \pi n_i \frac{\nabla E^2}{4 \omega_0^2} 
\end{equation}
Introducing the maximum plasma density (achieved at the end of the laser pulse) $n_i^{(max)}$ , $\omega_p^{max}=\sqrt{4 \pi n_i^{max}}$, $\theta=t \omega_p^{(max)}$, $D=\tau \omega_p^{(max)}$, $A=\frac{E_0^2 \omega_p^{(max)}}{c \omega_0^2 }$, and taking into account that 
$n_i (\theta)=\int_{-\infty}^{t=\theta/\omega_p^{max}} w(t') dt' = \frac{1}{\omega_p^{max}}\int_{-\infty}^{\theta} w(\theta) d \theta $
we arrive at
%
\begin{equation}
\begin{array}{cc}
     &  \partial^2_{\theta \theta}E_s+ \frac{n_i(\theta)}{n_i^{(max)}} E_s + \left(\frac{\nu_c}{\omega_p^{(max)}} - \frac{w(\theta)}
{\int_{-\infty}^{\theta} w(\theta') d \theta'}
\right) \partial_\theta E_s=\\
     & - A \frac{n_i(\theta)}{n_i^{(max)}} \frac{\theta } {D^2} \exp{\left(-\frac{2 \theta^2}{D^2}\right)}.
    \label{dthth_E_s1}
\end{array}
\end{equation}
From this Eq. one can see that under $\nu_{c} \ll \omega_p^{(max)}$ the result ($E_s/A$) does not depend either on the medium density, or on the intensity-independent factor in $w$ (these factors appear both in the numerator and denominator, so that they cancel).

Starting from Eq.~(\ref{dtt_E_s}) the equations were derived for $x=0$. As one can see from Eq.~(\ref{dtt_v}) the dependence of $E_s$ and the plasma properties on $x$ is defined by this dependence for $\Pi(x,t)$. So the plasma wave propagates in the medium with the laser pulse velocity. Taking into account Eq.~(\ref{pulse}) we have $\partial_x E_s= (-1/c) \partial_t E_s$. Then from Eq.~(\ref{d_E_s}) we find:
\begin{equation}
    \rho=\partial_\theta E_s\frac{\omega_p^{(max)}}{c 4 \pi}.
\end{equation}
The latter Eq. can be rewritten as:
\begin{equation}
    \rho=B \ \partial_\theta E_s/A,
\end{equation}
where $$ B = n_i^{(max)}\frac{E_0^2}{c^2 \omega_0^2}.$$

\subsection{Results}

\begin{figure}
a)

\includegraphics[width=0.45\textwidth]{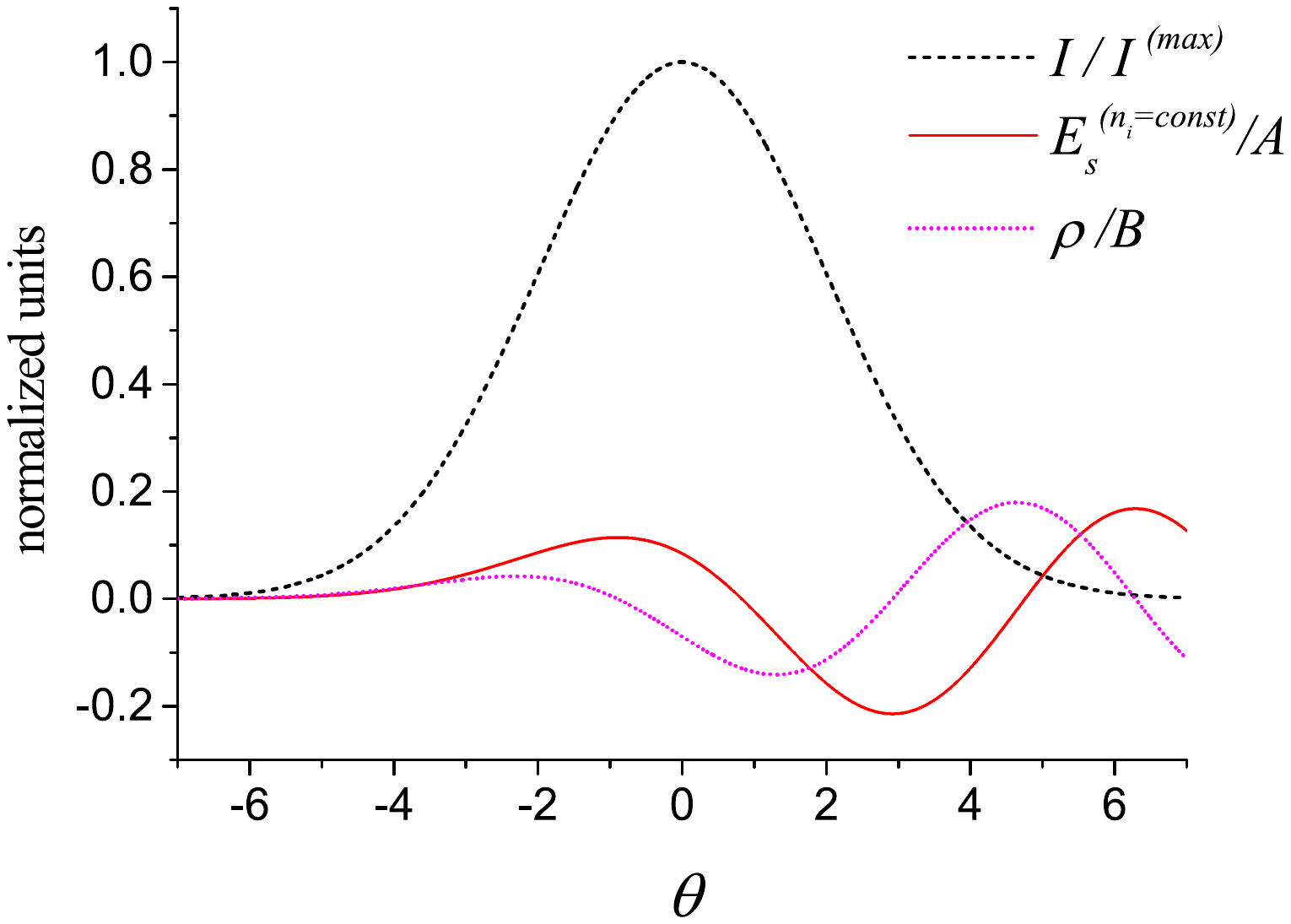}

b)

\includegraphics[width=0.45\textwidth]{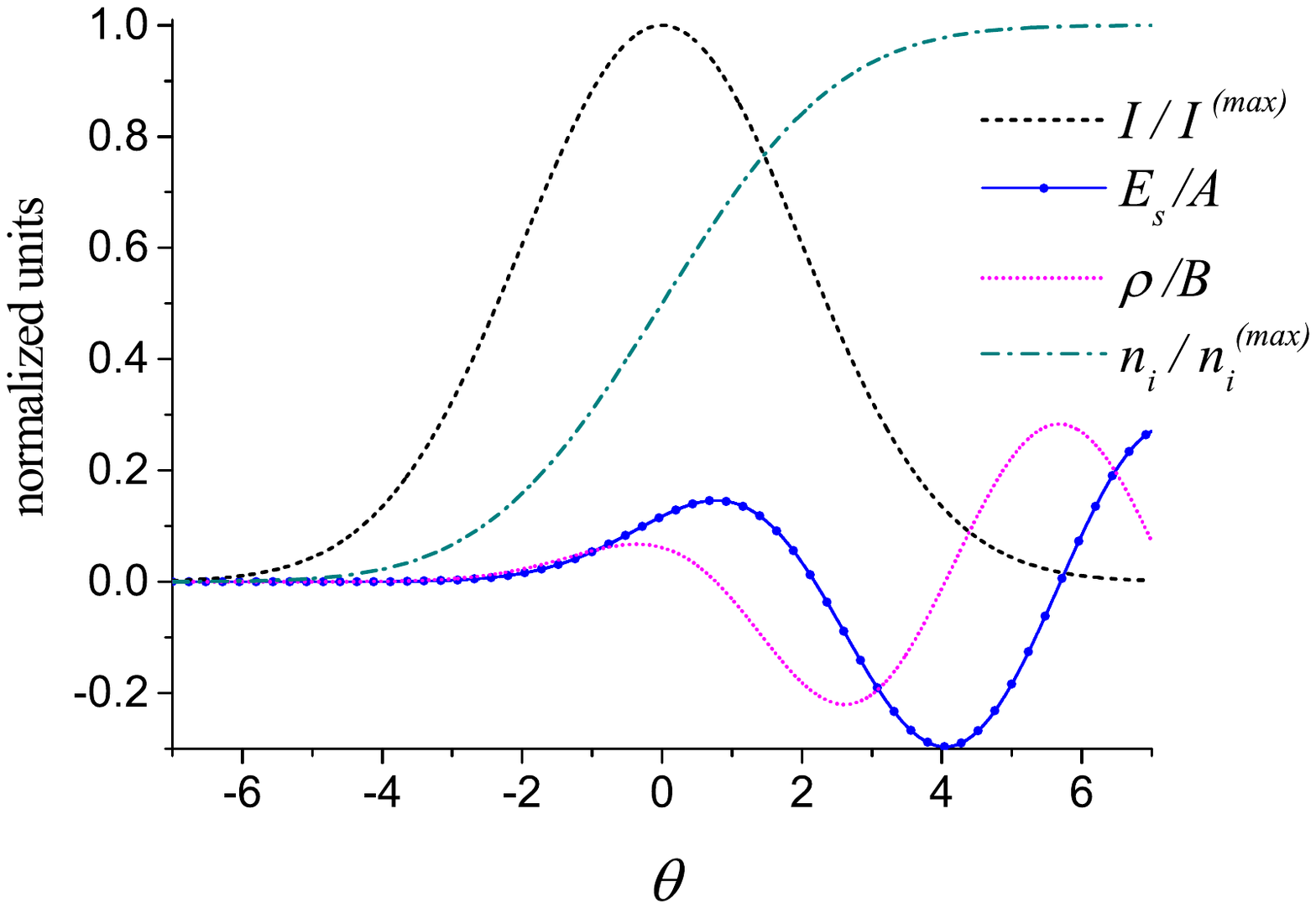}

\caption{The normalized plasma field $E_s/A$ and the charge density $\rho/B$ for $n_i=const$ (panel a) and $n_i \ne const$ (panel b))  calculated for $D=4$ as functions of $\theta$. The normalized laser intensity (dashed black curve), the normalized plasma field (solid red and blue curves),  the normalized charge density (dotted magenta curve), and the normalized plasma density (dash-dotted cyan curve).}
\label{fig:E_s}
\end{figure}

Fig.~\ref{fig:E_s} shows $E_s(x=0, \theta)/A$, $\rho(x=0, \theta)/B$ and $n_i(x=0, \theta)/n_i^{(max)}$ calculated numerically for $D=4$ and $\nu_{c} \ll \omega_p^{(max)}$ using Eqs.~(\ref{dthth_E_s}) and~(\ref{dthth_E_s1}). The calculated results vary but slightly with a change in the $w(I)$ dependence. We used $w(I) \sim I$; such dependence takes place for \textcolor{black}{the laser intensities close to those we are using here}  (see~\cite{Bauer_1999} for more details).  One can see that the maximum value of $E_s$, achieved within the laser pulse (denoted below as $E_s^{(max)}$) is reached near the maximum of the pulse.

\begin{figure}
\includegraphics[width=0.45\textwidth]{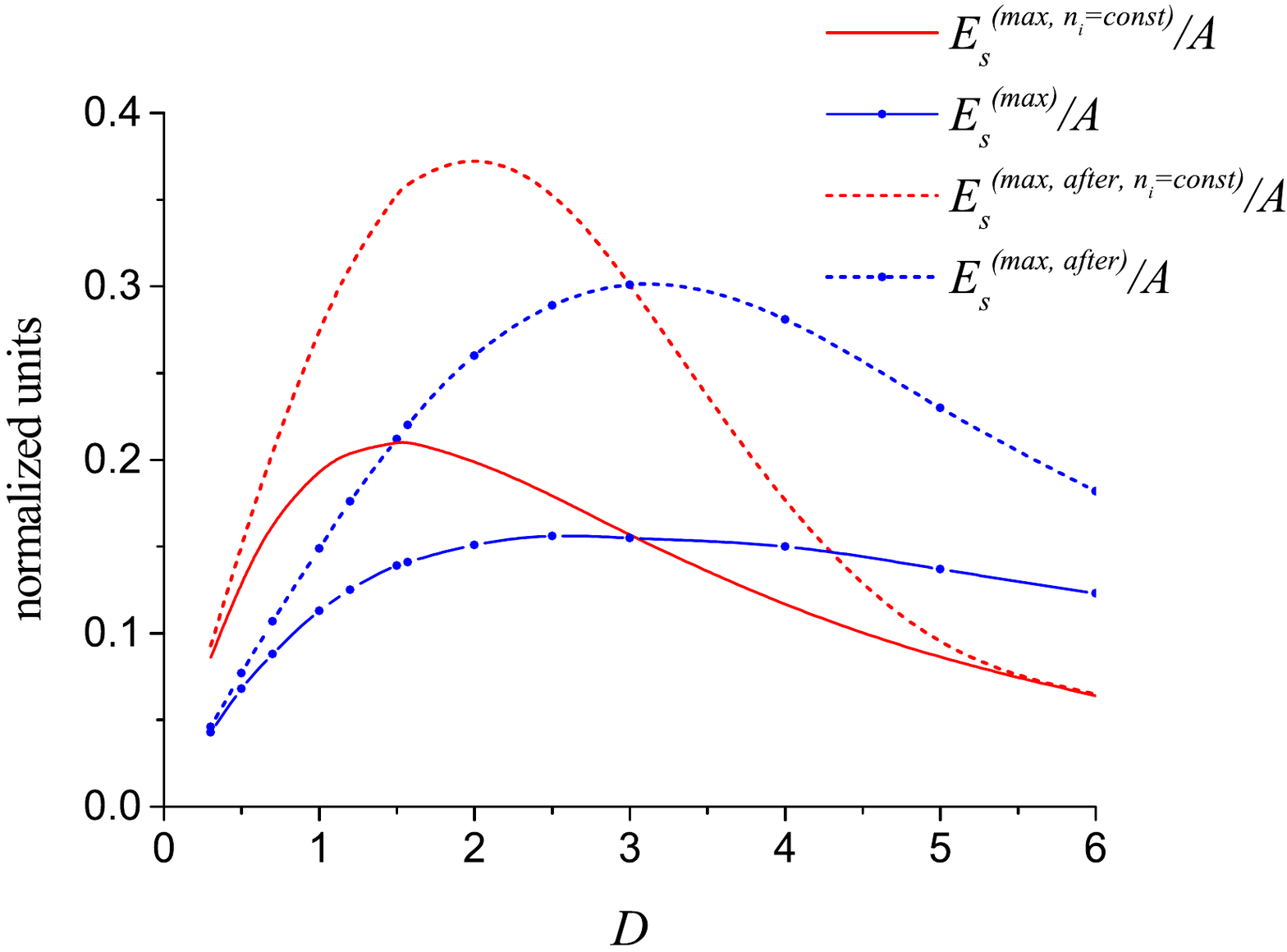}
\caption{The normalized maximum plasma field achieved within the laser pulse (solid curves) and the amplitude of its oscillations after the pulse (dashed curves) as functions of $D$. Results for $n_i=const$  (red curves) and for $n_i \ne const$ (blue curves with circles). Note that the normalized amplitude of the charge density oscillations equals that of the plasma field oscillations. }
\label{fig:E_s^max}
\end{figure}

Fig.~\ref{fig:E_s^max} shows $E_s^{(max)}/A$ as a function of $D$ (under  $\nu \ll \omega_p$). One can see that for the case of $n_i=const$ the maximum value $E_s^{(max)}/A=0.211$ is achieved for $D=\pi/2$. For the case $n_i \ne const$ the maximum value is smaller ($E_s^{(max)}/A=0.157$). It is achieved for $D \approx 3$. The figure shows also the normalized oscillation amplitudes of the electrical field $E_s^{(max, after)}/A$ and the charge density $\rho^{(max,after)}/B$ after the pulse (the latter two normalized amplitudes are equal).

The maximum of $E_s^{(max, after)}/A=0.38$ for $n_i=const$ is achieved for $D=2$. Note that in~\cite{Sm10} considering a circularly polarized laser field they have found that the maximum is twice higher than the one we have found in the case of linear polarization (this is the usual difference in the plasma wave excitation efficiency in circularly- and linearly-polarized fields, see Review~\cite{Sm4}) and the maximum is achieved for the same value of $D$. 

\begin{figure}
\includegraphics[width=0.45\textwidth]{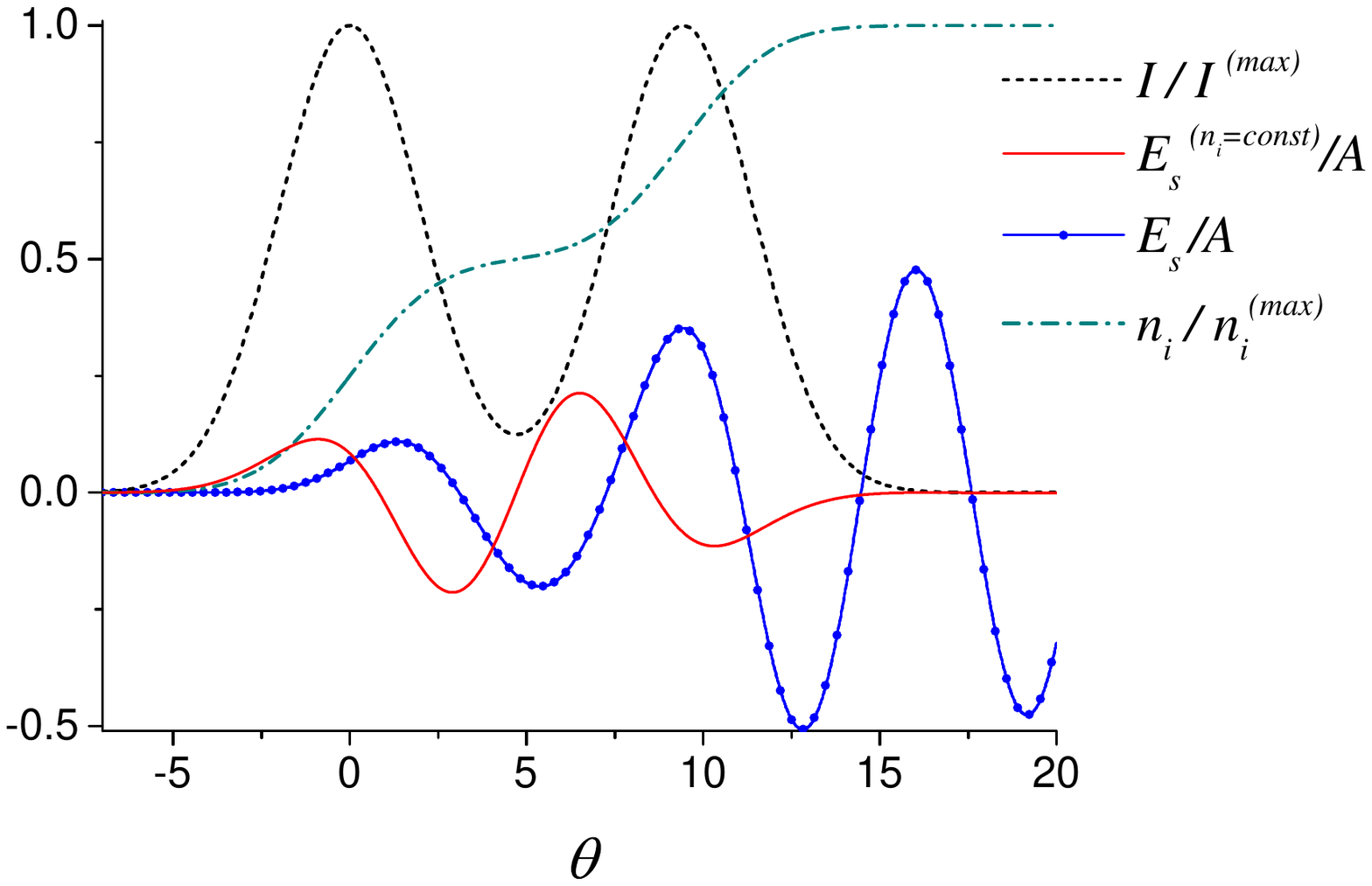}
\caption{The normalized plasma field $E_s/A$ for $n_i=const$ (red curve) and $n_i \ne const$ (blue curve with circles)  excited by two pulses with equal durations of $D=4$ and the delay between the pulses of $1.5 \times 2 \pi / \omega_{p}$. }
\label{fig:E_s_two_pulses}
\end{figure}

The plasma wave can be generated more efficiently with a combination of laser pulses. The simplest case of two pulses with equal durations and intensities is shown in Fig.~\ref{fig:E_s_two_pulses}. For the chosen delay of $1.5 \times 2 \pi / \omega_{p}$ for the case $n_i=const$ the plasma wave after the second pulse is completely suppressed; for the case of $n_i \ne const$ the maximum normalized amplitude of the plasma field inside the (second) laser pulse is $E_s^{(max)}/A=0.35$ and after its passing it is $E_s^{(max, after)}/A=0.51$. 

\section{\label{HHG} Compensation of the recolliding electron's magnetic drift by the plasma field}
From Figs.~\ref{fig:E_s},~\ref{fig:E_s^max} and from Eq.~(\ref{A}) one can see that for the parameters of the laser field typical for studies of the rescattering processes (for instance, HHG), the plasma wave field is much weaker than the laser field (in more details, $E_0/\omega_0 \sim 1$ in the atomic units, $c=137$, $\omega_p<\omega_0<<1$ ). However, its role in the rescattering can be non-negligible for certain conditions as described below.

\begin{figure}
\includegraphics[width=0.35\textwidth]{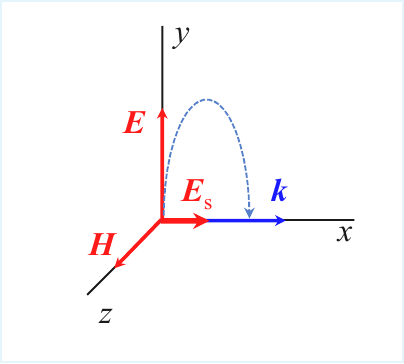}
\caption{The fields of the the electromagnetic wave $\textbf{E}$ and $\textbf{H}$, its wave vector $\textbf{k}$, the electric field of the plasma wave $\textbf{E}_s$. The dotted curve shows the electron's trajectory before rescattering. The Lorentz force accelerates the electron in the pulse propagation direction and the electric field of the plasma wave accelerates it in the opposite one.}
\label{fig:illustration}
\end{figure}

\textcolor{black}{During photoionization the linear momentum of the absorbed photons are partially transferred to the photoelectrons
~\cite{PhysRevLett.106.193002, PhysRevLett.113.263005, PhysRevLett.113.243001, PhysRevLett.118.163203, NatPh.15.1222, PhysRevLett.123.093201}, in particular, leading to non-zero average longitudinal velocity of the photoelectrons. Their longitudinal momentum distribution was recently studied both theoretically and experimentally in~\cite{NatCommun.10.5548, PhysRevLett.125.073202, PhysRevA.105.L031102, PhysRevA.105.053107, PhysRevA.110.063113}.} For high intensities and/or low-frequencies of the laser pulse the longitudinal drift of the photoelectron affects the rescattering, see Fig.~\ref{fig:illustration}. Namely, after the detachment of an electron from the parent ion the Lorentz force accelerates the electron in the $x$-direction (the direction of the laser pulse propagation); so the electron's wave packet spreads (due to quantum-mechanical uncertainty) and drifts in the propagation direction. If at the recombination instant this magnetic drift exceeds the wave-packet's size, rescattering is \textcolor{black}{suppressed~\cite{Taranukhin_2000,Walser_2000,Emelin_2013, Emelina_2014, Emelina_2015, PhysRevLett.118.093001}. Moreover, even in case of moderate drift the electron momentum distribution in NSDI and high-order ATI is strongly affected~\cite{Agaphy} and the magnetic field effect can be used as a tool to study electrons' correlation~\cite{PhysRevLett.128.113201}.} 

Let us consider the rescattering near the top of the laser pulse. The Lorentz force moves the electron's wave-packet in the pulse propagation direction, and the field of the plasma wave moves it in the opposite one. (Quantitatively this can be understood as follows: at the leading edge of the pulse the free electrons are accelerated forward, so the electron density at the leading edge is {\it higher} than the ion one. This cloud of electrons repulses the free electron that appears due to photoionization near the top of the laser pulse.) So, the action of the plasma wave can compensate the magnetic drift.

\textcolor{black}{
Consider the quasi-classic trajectory~\cite{3-step_C} of an electron which has been detached from the parent ion at the time $t_i$, accelerated by the laser field $F \sim \cos(\omega_0 t)$, and returned back to the origin at the time $t_r$. These times are defined by the parameter $\varphi$ ~\cite{Platonenko_review}:
$$
t_i=\frac{1}{\omega_0}\arctan{\frac{1-cos \varphi}{\varphi-sin \varphi}},
$$
$$
\varphi= \omega_0 \tau_{free},
$$
where
$$
\tau_{free}=t_r-t_i
$$
is the time of the electron's free motion. This time (or $\varphi$) defines the kinetic energy of the returning electron~\cite{3-step_C}:
$$
\varepsilon(\varphi)= U_p\frac{(2-2\cos \varphi -\varphi \sin \varphi)^2}{1-\cos \varphi-\varphi \sin \varphi+\varphi^2/2}
$$
($U_p=\frac{E_0^2}{4 \omega_0^2}$ is the ponderomotive potential), and thereby the energy of the emitted XUV photon. The magnetic drift at the instant of the return is~\cite{Taranukhin_2000}:
\begin{equation}
    x_{Lor}=\frac{E_0^2}{\omega_0^3 c }  f
    \label{x_Lor}
\end{equation}
where 
\begin{equation}
\begin{array}{cc}
     & f=  -\sin(2 \omega_0 t_r)/8+ \sin(\omega_0 t_i) [\cos(\omega_0 t_r)-\cos(\omega_0 t_i)]+ \\
     &  \omega_0 \tau_{free}[\sin^2(\omega_0 t_i)+\cos(2 \omega_0 t_i)/4]+\sin(2 \omega_0 t_i)/8. 
\end{array}
  \label{f}
\end{equation}
}

\textcolor{black}{
On the other hand, the plasma field $E_s$ can be considered as a static one within the duration of the free motion of the electron's wave-packet before rescattering, so this field provides a shift of the wave-packet:
\begin{equation}
    x_{pl}=E_s \tau_{free}^2/2,
    \label{x_pl}
\end{equation} 
The strength of the plasma field $E_s$ is found using the results of Sec. I. Let us consider the plasma field $E_s^{(max)}$ close to its maximum achieved under the laser pulse duration $\tau=4/ \omega_p$ so that $D=4$, see Eq.~(\ref{D}), and $E_s^{(max)}\approx 0.14 A$, see Fig.~\ref{fig:E_s^max}. Defining $A$ from Eq.~(\ref{A}), we find from Eqs.~(\ref{x_Lor}) and (\ref{x_pl}):
\begin{equation}
\frac{x_{pl}}{x_{Lor}} = g(\varphi) \frac{\omega_p}{\omega_0},
    \label{Lor_pl}
\end{equation}
where
$$
g(\varphi) \approx 0.075 \varphi^2/f(\varphi)
$$
and $f(\varphi)$ is given by Eqs.~(\ref{f}). 
}

\textcolor{black}{
Thus, the magnetic drift can be compensated by the shift due to the plasma field, so that $x_{pl}/x_{Lor} \approx 1$. From Eq.~(\ref{Lor_pl}) we see that for different electron's trajectories this compensation takes place for different $\omega_{p}= \omega_0/g(\varphi)$.
}

\begin{figure}
\includegraphics[width=0.5\textwidth]{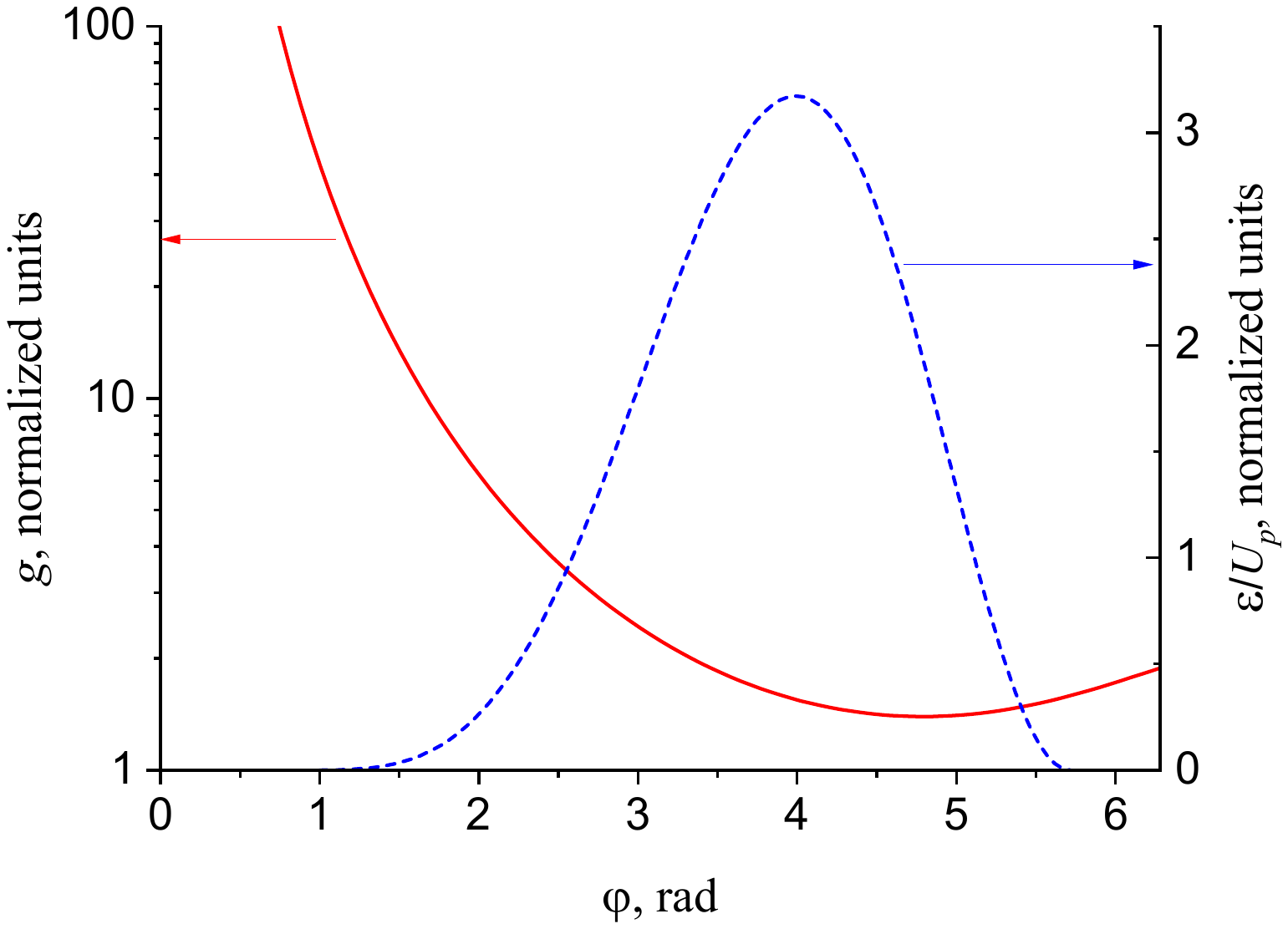}
\caption{\textcolor{black}{The coefficient $g$ and the energy of the returning electron $\varepsilon$ as functions of $\varphi=\omega_0 \tau_{free}$, calculated for the quasi-classical electron's trajectory (note the logarithmic scale along the left vertical axis and the linear scale along the right one). }}
\label{Lor_vs_pl}
\end{figure}

\textcolor{black}{
Fig~\ref{Lor_vs_pl} shows the coefficient $g$ together with the energy of the returning electron $\varepsilon$ as functions of $\varphi$.
From the figure we can see that for the cut-off trajectory (i.e. the one providing highest harmonics near the plateau cut-off, $\varphi\approx 4.1$) the compensation  takes place for a relatively high plasma density and low frequency of the laser field, so that $\omega_p \approx \omega_0/1.5$. Note that HHG can hardly be effective under such  high plasma density, leading to weak HHG phase-matching and degradation of the harmonic radiation coherence due to the plasma microfield~\cite{Strel_Plat_Bec_2005}. Moreover, the laser pulse duration should also be very short. Namely, $D = 4$ corresponds to the pulse duration (full-width at half-maximum of the intensity) $\tau_{FWHM}\approx 1.2 T_0$, where $T_0=2 \pi/\omega_0$ is the laser cycle duration. 
}

\textcolor{black}{
Moreover, from Fig.~\ref{Lor_vs_pl} one can see that for the "short" electron's trajectory ($\varphi < 4.1$) the compensation takes place for lower plasma density. Consequently, the required laser pulse duration can be longer (for instant, $D = 4$ under $\omega_p = \omega_0/10$ corresponds to $\tau_{FWHM}\approx 8 T_0$). Such values of the plasma density and the pulse duration are typical for HHG experiments. Note that this is the "short" trajectory that provides a major contribution to the experimentally observed HHG signal for the plateau harmonics because the signal due to the "long" one has much higher divergence. 
}

In the case of plasma wave excitation by a pair of pulses (Fig.~\ref{fig:E_s_two_pulses}) the plasma field is higher and \textcolor{black}{making a similar analytical calculation we find: $$\frac{x_{pl}}{x_{Lor}} \approx 2.3 \, g(\varphi) \frac{\omega_p}{\omega_0}, $$ so the compensation takes place for longer pulse durations and/or lower plasma densities.} Further relaxation of the requirements to the plasma density and/or pulse duration can be achieved using the first pulse with higher intensity and/or circular polarization. In such a configuration the first pulse generates the plasma wave and the rescattering process is studied in the second linearly-polarized pulse.

\textcolor{black}{
Thus, from the results of this section, one can conclude that the suppression of the HHG efficiency taking part at high laser intensities and low frequencies  due to the electron's magnetic drift can be eliminated compensating the drift by the shift of the electron by the plasma field. This compensation is hardly practical for the cut-off harmonics (at least in the simplest single-pulse experiment), but it is feasible for the plateau ones. Moreover, the efficiency of other rescattering processes (such as NSDI, high-order ATI and others) can be improved as a result of this compensation. In general, the plasma wave field can affect the angular distribution of the NSDI and high-order ATI photoelectrons and their correlation, similar to how the magnetic drift was shown to affect these properties in Ref.~\cite{Agaphy, PhysRevLett.128.113201}; the modification of the photon momentum transfer to the photoelectron~\cite{Mao_2025} due to the plasma wave could be another outlook of this study.}   

\section{\label{FID} Role of the plasma field in the XUV free induction decay}

To calculate the spectrum emitted via XFID in the presence of the field $E_s(t)$ we numerically solve the time-dependent Schr\"odinger equation (TDSE) 
\begin{equation}
i\frac{\partial}{\partial t}\Psi(\textbf{r},t) = \left(-\frac{1}{2}\nabla^{2}+V(\textbf{r})-x \, E_s(t)\right)  \Psi(\textbf{r},t)
\label{TDSE}
\end{equation}
for a model 2D atom in the single-active electron approximation. The~equation is solved in Cartesian coordinates: $\textbf{r}=\{x,y\}$. A~modified Coulomb potential~\cite{Strel_Plat_Bec_2005} providing the ionisation energy close to that of argon is used:
\begin{equation}
V(r)=  - \frac{1+A \exp(-\sqrt{r^2+b^2})}{\sqrt{r^2+a^2}} ,
\label{V}
\end{equation}
with $A = 2.8$, $a = 2.1$, $b=0.2$. See~\cite{Strel_Plat_Bec_2005, KhokhlovaStrelkov} and references therein for more details on the numerical method. 

The TDSE is solved at the time interval when the laser pulse has already passed, so Eq.~(\ref{TDSE}) contains only the field of the plasma wave $E_s(t)$. The effect of the laser field is taken into account via specific initial wave function $\Psi(\textbf{r},t=0)$. This function contains the ground state as well as the excited ones which can be populated via the laser field\footnote{\textcolor{black}{Another possible approach  for addressing this problem would be solving the TDSE} \textcolor{black}{ starting from the beginning of the laser pulse taking into account both the laser field and the field of the plasma wave. However, such approach requires huge numerical resources. First, one has to use a \textit{large} spatial numerical box with dimensions defined by the electron's excursion in the intense laser field , which is quite large on the atomic scale. Second, the temporal scale of the calculation should be sufficiently \textit{long} to take into account the plasma field existing during picoseconds. Our approach is less computationally expensive because we make long calculations but limit them to a moderate numerical box. Its size is large enough to accommodate the excited states; their spatial scale, however, is much smaller than the electron's excursion in the laser field. Our approach allows a correct description of an XFID line modification due to the plasma wave. To find the absolute line intensity or to describe the relative intensities of several lines, one should use the above-described direct approach.}} (polarized in the $y$-direction). Note that the excited states with different values of orbital momentum are populated because both single- and few-photon transitions from the ground state are possible.  

Note that in this section we do not take into account any relativistic effects (for instance, the electron wave-packet drift). Note also that the electric fields act in the $y$-direction (the laser pulse) and in the $x$-direction (the plasma wave field). Thus, there is no $z$-projection of the atomic polarization and the approximation of the real atom with a 2D system is reasonable.

Let us consider the radiation (emitted due to the XFID) propagating in the $x$-direction. The intensity of this radiation is proportional to the $y$-projection of the medium polarization squared. According to the Ehrenfest's theorem the polarization is proportional to the expectation value of the  force:
\begin{equation}
    f_y(t)=-<{\Psi(\textbf{r},t)}|{\frac{\partial}{\partial y} V(\textbf{r})}|{\Psi(\textbf{r},t)}> \, . 
\label{polarization}
\end{equation}

The TDSE is solved at the time interval of 1 ps.  Then the response~(\ref{polarization}) is multiplied by the temporal mask: $\tilde f_y(t)= f_y(t) \exp(-t/T_2)$ taking into account the loss of the emission's coherence. We assume that the decoherence time (also known as transverse relaxation time, see Ref.~\cite{Shen}) $T_2$ is 300 fs.

\begin{figure}
\includegraphics[width=0.5\textwidth]{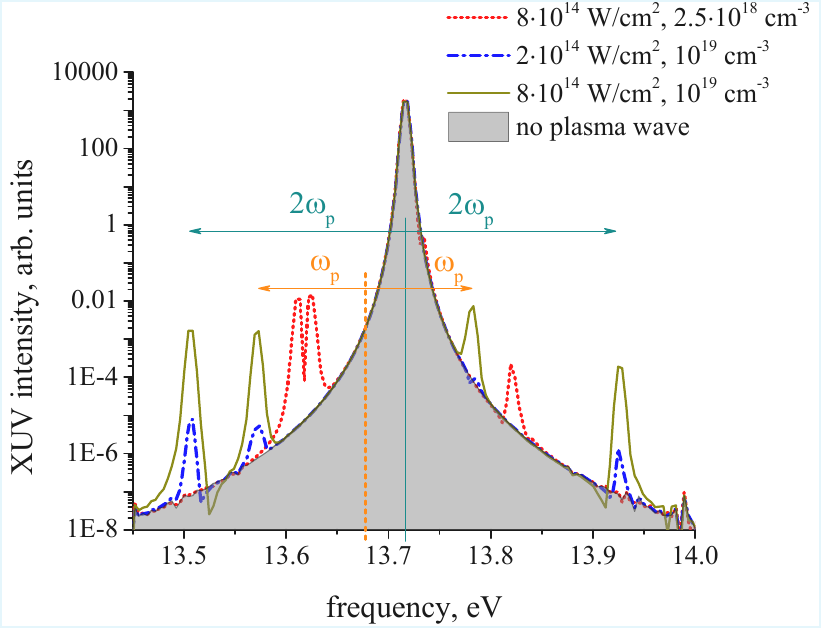}
\caption{The XUV free induction decay spectra \textcolor{black}{(note the logarithmic scale along the vertical axis)} emitted in the presence and in the absence of the plasma wave near the frequencies of the allowed (13.72 eV, solid magenta line) and forbidden (13.68 eV, dashed orange line) transitions. The peak laser intensities and the electron densities are shown in the graph, \textcolor{black}{the laser wavelength is 800 nm}, $\omega_p$ is the plasma frequency for the electron density of $10^{19}$~cm$^{-3}$. \textcolor{black}{The intensities of the plasma wave are found for the laser intensities and the plasma densities indicated in the legend, using the results of Sec. I, see text for more details}}
\label{fig:xFID}
\end{figure}

Figure~\ref{fig:xFID} shows the XUV spectra ($|\tilde f_y(\omega)|^2$) emitted in the presence and in the absence of the plasma wave. The electron density of $2.5 \times 10^{18}$~cm$^{-3}$ corresponds to the plasma frequency $0.054$~eV and the density of $10^{19}$~cm$^{-3}$ corresponds to $0.108$~eV;  the intensity of the plasma wave is $8.2 \times 10^{6}$~W/cm$^{2}$ for the dash-dotted blue curve, $3.3 \times 10^{7}$~W/cm$^{2}$ for the dotted red one, $1.3 \times 10^{8}$~W/cm$^{2}$ for the solid yellow one. \textcolor{black}{These intensities of the plasma wave are found using results of Sec. I for the laser and plasma parameters presented in the figure. Namely, as in Sec. II, we assume the laser pulse duration $\tau=4/ \omega_p$  so that $D=4$, see Eq.~(\ref{D}); in contrast to Sec II, however, in this case we are interested in the plasma wave field {\it after} the laser pulse, so its strength is $E_s^{(max, after)}\approx 0.28 A$, see Fig.~\ref{fig:E_s^max}. We calculate $A$ using Eq.~(\ref{A}) for the laser intensities and the plasma densities presented in Figure~\ref{fig:xFID}, then calculate the strength and the intensity of the plasma wave.}

One can see that for the laser intensity and the medium density typical for the XFID experiment the presence of the plasma wave leads to weak but detectable peaks in the emitted spectra. In the vicinity of the allowed transition (solid cyan line) these peaks are the satellites shifted by $\pm 2 \omega_p$ from the strong XFID line emitted in the absence of the plasma wave (i.e. under low medium density). For the case of the forbidden transition (dotted orange line) the satellites are shifted by $\pm \omega_p$ from the transition frequency. So, our calculations show that observing the XFID spectrum in the presence of the plasma wave allows measuring of the forbidden transitions' frequencies. 

\section*{Conclusions}

In this paper, the role of the electric field of the plasma wave on the rescattering processes is studied analytically and numerically. In the 1D approximation we derive analytical equations describing the amplitude of the plasma wave as a function of the laser pulse intensity, frequency, and duration, as well as of the plasma density. Using this result, we find conditions in which the field of the plasma wave can compensate for the rescattering electron's magnetic drift, which otherwise dramatically suppresses the rescattering efficiency in intense low-frequency laser fields. \textcolor{black}{Thus the efficiency of such rescattering processes as NSDI or recollisional excitation can be improved under proper plasma density, providing this compensation.  For the  cut-off HHG the required plasma density is relatively high which might prevent the increase of the overall generation efficiency. However, for the plateau harmonics the required plasma density is closer to its typical experimental values thus the efficiency increase due to the compensation of the magnetic drift is feasible.} Finally, solving numerically the TDSE for an atom emitting XUV due to free-induction decay, we simulate the XFID process in the presence of the plasma wave. We show that this field leads to the appearance of the satellite lines near the allowed transition frequencies and also to the appearance of pairs of new lines shifted by $\pm$plasma frequency from the forbidden transition frequencies. Thus, observation of these lines allows the detection of the forbidden transition frequencies via the XFID spectroscopy. 

\section*{Acknowledgment}
This study was funded by RSF through Grant No. 24-22-00108.

\bibliography{lit}

\end{document}